\documentclass[10pt, twocolumn]{IEEEtran}
\usepackage{ifpdf} %% 切换latex 和pdflatex 命令编译
\usepackage{array} %%% 排列包（创建排列，1、2、3 这种排列）

\usepackage{amsmath} %% 数学包
\usepackage{color}

\usepackage{amsthm} % 如果要使用proof 语句就必须要引入这个包

\theoremstyle{remark}

\theoremstyle{remark}
\usepackage{amssymb}    %
\usepackage{amsfonts}
\usepackage[english]{babel}
\usepackage{cite}
\usepackage{multirow}
\usepackage{multirow}
\usepackage{makecell}
\usepackage{stfloats}    % big equations float
\usepackage{algorithm}   %format of the algorithm
\usepackage{algorithmic} %format of the algorithm
\usepackage{graphicx}
\usepackage{caption}
\usepackage{epsfig}
\usepackage{subfigure} %subfigs
\usepackage{cite}
\usepackage{tablefootnote}
\usepackage{diagbox}   % 表格内斜线

\usepackage{setspace} %for begin{spacing} for complex tables
\usepackage{threeparttable}
\usepackage{upgreek}
\usepackage{url}

\usepackage{footnote}
\usepackage[textsize=tiny]{todonotes}
\usepackage{marginnote}
\usepackage{pifont}
\usepackage[perpage]{footmisc}  % 每页脚注重新编号

\usepackage{booktabs}    % 表格

\allowdisplaybreaks  %allow equation crossing pages

 \usepackage[draft]{changes}   % 审阅模式 % draft or final

\usepackage{color}

\begin{document}
\title{How Far Are Wireless Networks from Being Truly Deterministic?}


\author{Yan Li, Yunquan~Dong,~\IEEEmembership{Member,~IEEE},
                  Pingyi Fan~\IEEEmembership{Senior Member,~IEEE}
                 and Khaled~Ben~Letaief,~\IEEEmembership{Fellow,~IEEE}

        \thanks{Y. Li and Y. Dong are with the School of Electronic and Information Engineering,  Nanjing University of Information Science and Technology, Nanjing 210044, China  (e-mail: \{yanli, yunquandong\}@nuist.edu.cn).

        %P. Fan is with the Beijing National Research Center for Information Science and Technology(BNRist), Beijing 100084, China.
            P. Fan is  with the Department of Electronic Engineering, Tsinghua University, Beijing 100084, China (e-mail:  fpy@tsinghua.edu.cn).

       Khaled B. Letaief is with the Department of Electrical and Computer Engineering, HKUST, Clear Water Bay, Kowloon, Hong Kong (e-mail: eekhaled@ust.hk).

       % The work of Y. Dong was supported by the National Natural Science Foundation of China (NSFC) under Grant 61701247, the open research fund of National Mobile Communications Research Laboratory, Southeast University, under grant No. 2020D09, and the Startup Foundation for Introducing Talent of NUIST under Grant 2243141701008.
%            The work of P. Fan was supported by the China Major State Basic Research Development Program (973 Program) No. 2012CB316100(2) and Beijing Natural Science Foundation No. 4202030.
        }
        }

\maketitle

\vspace{-20mm}

\begin{abstract}
    With the rapid development of Internet-of-Things (IoT) technology and machine-type communications, various emerging applications appear in industrial productions and our daily lives.
 Among these, applications like industrial sensing and controlling, remote surgery, and automatic driving require an extremely low latency and a very small jitter.
    Delivering information deterministically has become one of the the biggest challenges for modern wire-line and wireless communications.
 In this paper, we present a review of currently available wire-line deterministic networks and discuss the main challenges to build wireless deterministic networks.
    We also discuss and propose several potential techniques  enabling wireless networks to provide deterministic communications.
 By elaborating the coding/modulation schemes of the physical layer and managing the channel-access/packet-scheduling at the media access control (MAC) layer, it is believed that wireless deterministic communications can be realized in the near future.
\end{abstract}


\IEEEpeerreviewmaketitle

\section{Introduction}
\IEEEPARstart{W}{ith} the rapid development of the Internet of Things (IoT) and Industry 4.0, an increasingly large number of smart devices are connected through the internet.
    This has spawned a lot of time-critical applications, such as the smart grid, the remote control and the factory automation \cite{A.N}.
Most of these time-critical applications require the smart devices to complete information deliveries with deterministic delays, which is critical to the system performance but difficult to achieve in practice.
    Note that traditional wire-line and wireless networks (e.g., Ethernet) only provide best-effort transmissions, in which the end-to-end latency may be as high as 50 $\sim$ 500 ms and the packet loss rate may be up to 5\% $\sim$ 50\%\cite{T.H}.
    In order to support information transmissions with low latency, low jitter, and high reliability, therefore, many deterministic networking technologies have been developed \cite{T.H}.

The deterministic property of networks can be interpreted in the following two ways.
    First, it is widely recognized that in a deterministic network, the packets should be delivered with \textit{small delays} (e.g., end-to-end delay being smaller than 10 ms) and \textit{near-zero jitters} (approximately several milliseconds)\cite{T.H}.
In a traditional soft real-time system, the packets are required to be received before a deadline with few violations. In a hard real-time system, the packets must be received before the deadline without any violations. In contrast, a deterministic network tries to complete each packet delivery at its scheduled deadline with near-zero divergence.
    Second, there are also deterministic networks where the packets are expected to be received \textit{exactly at some planned epochs}, i.e., be received on-time\cite{On-time-LY}.
In a nutshell, these deterministic networks concentrate on the \textit{predictability} more than the immediacy of transmissions.
    Thanks to the ability to provide bounded end-to-end latency and jitter, the deterministic networks have a wide range of applications in various areas of industrial automation.
A few examples include:
\begin{itemize}
      \item \textit{real-time monitoring systems} like programmable logic controller (PLC) controlled systems and unmanned logistics systems have a strong demand on the real-time status of the operational processes and equipment status within the monitoring field.
            Specifically, thousands of field sensors transmit their sensing information on the equipment status and task execution progress to the control center, so that it can control actuators, initiate new control tasks, arrange maintenance or trigger alarms in an automatic or human intervention manner.
          To guarantee the timeliness and reliability of the received data, a 1$\sim$10 ms deterministic delay is required for the data exchanging among the sensors, actuators, and the control center\cite{5gdnwhite2020}.
      \item \textit{in-vehicle networks}  should provide real-time and reliability assurance for the transmission of braking/steering commands and advanced driver assistance systems (ADAS) type data for vehicles.
      Thus, a series of deterministic networking mechanisms (e.g., the controller area network-bus (CANbus), automotive Ethernet and audio video bridging (AVB) system) were developed to meet the strict requirement on transmission delay and delay jitter\cite{H.G}.
      %fast increase in the network bandwidth and the flexility of network expansion. % \cite{R} \cite{W}.
      \item \textit{precise space-time positioning system}, in which the satellites provide precise location and timing for ground users with errors less than 1 cm and 1 $\upmu$s, is a key enabler of real-time digital twins and meta-verse\cite{J.Y}.
            Note that as a bridge between the physical world and a digital world, the virtual representations of the objects, events, and actions are  expected to follow the same spatial and temporal statistics as their physical counterpart.
          Thus, the positioning for the physical world is meaningful only if the transmission of corresponding timing information is precise and deterministic.
\end{itemize}

%Moreover, the controlling in power systems, power electronics systems, networked robot systems raised strict real-time requirements for information transmissions \cite{H.-D, C.L, L.Wang}.
Although deterministic networks have been widely used, most practical applications are based on wire-line networking technologies (e.g., the Ethernet, Profinet, EtherCAT, Powerlink) other than wireless ones.
    This is because the randomness in wireless channels has prevented the wireless technologies such as Bluetooth (IEEE 802.15.1), ZigBee (IEEE 802.15.4), Wi-Fi (IEEE 802.11) and cellular networks (3GPP) from providing the required low latency and high reliability.
For example, super-critical industrial control applications require communications networks to provide 100 $\upmu$s cycle times for dozens of nodes with extremely limited jitter\cite{M.Z}.
While real-time Ethernet can meet this requirement, the most advanced wireless solutions are hardly able to provide a cycle time on the order of 1 ms.
However, wireless networks are much cheaper to build and more flexible to changes.
    In this paper, therefore, we shall survey the existing wire-line solutions of deterministic networks, discuss the main challenges of replacing wire-lines with wireless channels, and propose several possible techniques enhancing the determinacy of wireless communications.

This rest of the paper is organized as follows.
    In Section \ref{Deterministic Networking Technologies}, we discuss some existing solutions to deterministic communications.
    In Section \ref{challenge}, we analyze the difficulties and challenges in providing deterministic communications through wireless based techniques.
    In Section \ref{Enabling deterministic}, we propose several physical layer and medium access control (MAC) layer solutions to ensure deterministic transmissions over wireless networks.
    Finally, we conclude the paper in Section \ref{Conclusion}.

\section{Existing Deterministic Networking Technologies}\label{Deterministic Networking Technologies}
The objective of deterministic networks is to provide deterministic latency, low jitter, low packet loss,  and high reliability for information delivery.
    Although the wire-line networks have relatively stable transmission capabilities, the corresponding end-to-end delays also depend on the burstiness of the source traffic.
For example, many bursts of data may arrive at a transmitter within a short period even if the overall traffic load is low.
    In this case, the output link would be congested, resulting in large latencies and a high packet loss rate.
That is, traditional Ethernet networks cannot provide quality of service (QoS) guarantees for real-time and deterministic applications without using some advanced high-layer scheduling.
    To this end, several deterministic networking technologies have been developed, including the time-sensitive networking (TSN) \cite{Time-sensitive.networking.2018}, the deterministic networking (DetNet) \cite{DetNet-2019}, and the deterministic Wi-Fi (Det-WiFi) \cite{DetWifi-2017}.

Specifically, TSN enhances the deterministic assurance of networks mainly through MAC layer techniques, such as the precise network synchronization (IEEE 802. 1 AS), shaping mixed traffic of different priorities, the flow reservation mechanism (IEEE 802. 1 Qcc), and time-sensitive traffic controlling mechanisms (IEEE 802. 1 Qch, cyclic queuing) \cite{Time-sensitive.networking.2018}.
    By eliminating micro-bursts and providing deterministic QoS guarantees, TSN has become the basis of time sensitive applications like industrial automation and automotive driving.

 The Internet Engineering Task Force (IETF)  proposed a DetNet system to provide ultra-low end-to-end latency and high reliability services through an optimized network layer routing scheme. Particularly, DetNet enables the co-networking of those real-time traffic with some less time-sensitive traffic flows.
    Key features of DetNet include explicit routing, multi-path routing, resource reservation, jitter reduction, flow replication/merging, packet sequencing, and so on \cite{DetNet-2019}.
However, most of the proposed DetNet standards draw their basis on TSN and suffer from difficulties in strict time synchronization, long transmission delays, and high complexity of traffic scheduling.
    Thus, practical implementations of DetNets need further investigations and improvements.

The IEEE 802.11be task group (TGbe) has developed a series of protocols to provide deterministic services over  wireless access networks, focusing on  PHY and MAC layer techniques and amendments \cite{CLC-11111}.
    In particular, 802.11 be has integrated multiple solutions, such as the access to the medium, scheduling/premption of packets, and the coordination among multiple access points.
 In this way, 802.11be is expected to be effective in reducing the worst-case end-to-end delay of wireless local area networks (WLANs) at a peak throughput of 30 Gbps.
    Thus, 802.11be is considered as the successor of the IEEE 802.11ax and a key element of next generation Wi-Fi, i.e., Wi-Fi 7\cite{CLC-2021}.
There were also works running software TDMA MAC over the high-speed 801.11 physical layer, which is named as Det-WiFi \cite{DetWifi-2017}. As shown in the paper, Det-WiFi support high-speed applications and provide better deterministic services in practical multi-hop industrial environments.

%\textbf{Currently deterministic network technologies cannot be applied to applications with strict real-time performance requirements such as power system control \cite{H.-D}, power electronics control \cite{C.L} and high-end motion control \cite{L.Wang} (IEEE 802.11 sends a single packet for at least 30us, 5G URLLC can not reduce the end-to-end delay to less than several hundred microseconds), so the solution for this type of applications is to use wire-line real-time Ethernet, such as Profinet, EtherCAT, Powerlink.}

\section{Main Challenges in Moving from Wire-line to Wireless Deterministic Networking}\label{challenge}
To meet the required ultra-low latency and ultra-high reliability in time-critical applications, the vast majority of communications technologies in industrial IoT  are wire-line based.
    However, the wire-line networks are less flexible and more expensive to build, as well as to maintain for long-term reliability.
Therefore, wireless networking is expected to be used to improve the flexibility of industrial productions.
    Nevertheless, wireless communications are inherently less reliable and less efficient than wire-line communications for the following reasons.

\begin{itemize}
      \item Susceptibility to external interference and multi-path fading. Due to the shared nature of wireless channels, each transmitted packet suffers from the interference from other users and adjacent frequency bands.
                Thus, wireless channels are prone to errors, which further leads to packet loss.
          In addition, the frequency selective property and the multi-path propagation property reduce the reliability of wireless transmissions significantly.
            This further increases the corresponding end-to-end delays and jitters.
      \item The half-duplex constraint limits efficiency. For wireless transmissions, most of the nodes are not allowed to transmit and receive simultaneously in the same frequency band, since the transmitted signals will cause strong self-interference to the received signals at the same node. Under the half-duplex constraint, the efficiency of the wireless channels are quite limited, and thus increasing the latency of transmissions.
          Although some preliminary results on full-duplexing have been available in wireless systems, the cost of full-duplex devices is relatively high and the size of full-duplex devices is large. Thus, the full-duplex technology is not mature for industrial environments at present \cite{Full-duplex-2010,  Full-duplex-2014}.
            %Under the half-duplex constraint, the efficiency of the wireless channels are quite limited, and thus increasing the latency of transmissions.
      \item Limited packet length.  In industrial Internet of Things (IIoT) networks, the packets exchanged among smart devices are usually short, with a payload size of a few bytes.
            This brings great difficulty for the channel coding and frame design for wireless transmissions.
        Specifically, the control overhead may have a greater impact on the delay than the payload itself.
                The transmission rate of short packets is also very limited due to the short coding block length.
\end{itemize}

Moreover, the transmitted packets are easily eavesdropped during transmissions due to the broadcasting nature of wireless channels.
     The scarcity in transmit power and frequency bandwidth also bring great challenges to the protocol design  for wireless networks.

\section{Enabling Determinacy by Wireless Networks}\label{Enabling deterministic}
\begin{figure}
  \centering
  \includegraphics[width=3.5in]{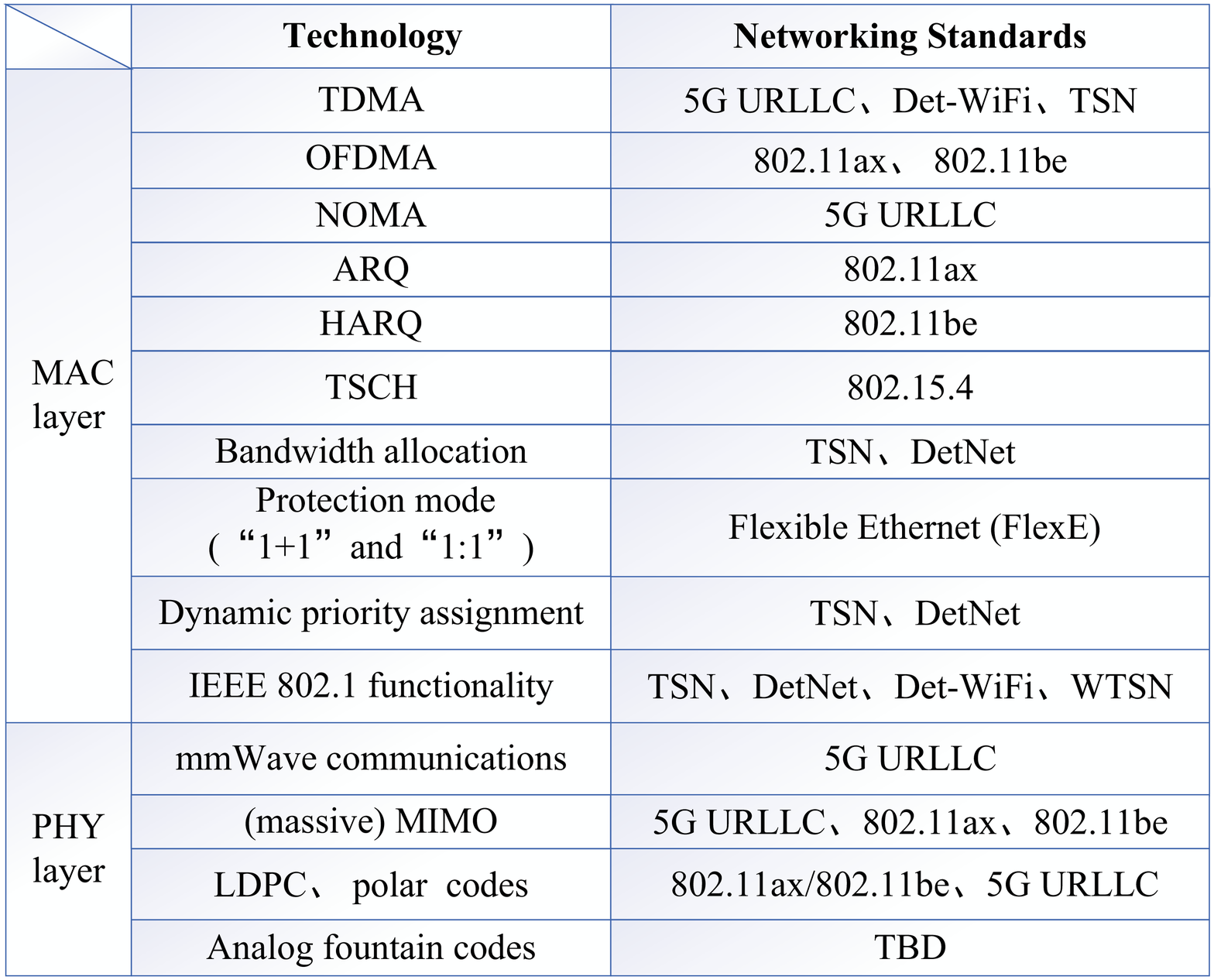}\\
  \caption{Technologies that have been or can be used in time sensitive wire-line/wireless networking standards.}\label{table_1}
\end{figure}

Due to the fading property and the attenuation property, the information transmissions over wireless channels are unreliable and rate-varying, i.e., indeterministic.
    For wireless channels, the key solution to provide deterministic transmissions lies in compensating the randomness of channels and shaping the source traffic flows according to the time-varying transmission capacity of the channels.
 %The key solution to provide deterministic transmissions lies in compensating the  randomness of channels and matching the source traffic with the channel.
  %For example, we can improve the reliability of each link and match changes in channel transmission conditions by shaping and scheduling arriving traffic as well as replicating/eliminating some traffic.
    For example, we can increase the reliability of each link, shape the arriving traffic, and delay (or even eliminate) some flows.
Fig. \ref{table_1} presents some MAC/PHY layer techniques that have been or can be used in networks/wireless networking standards. These techniques can provide deterministic transmission for data by reducing latency, improving reliability, and exploiting diversity gain and bandwidth utilization. In this section,  we focus on these techniques and review some potential techniques for wireless deterministic networking.

\subsection{MAC Layer Techniques}
A major task of MAC layer protocols are to coordinate the access of nodes to the common communications channel and prevent conflicts among them.
    Due to the broadcasting nature of wireless channels, if two nodes are located in the same area and transmit at the same time  in the same frequency band, their signals will be added up and conflict each other at the receiver, which prevents the correct decoding of packets and increases transmission latency.
Thus, it is crucial for the MAC layer to adequately manage the conflicts among nodes and transmit real-time traffic with higher priority.

\subsubsection{TDMA}
%Traditional Ethernet mostly uses a bus topology or a logical star network structure, connecting all devices in other paths through a single wire-line cable, where the central node coordinates user access to the channel by distributing scheduling information. A typical solution is time division multiple access (TDMA), in which each node transmits in a dedicated time slot and the other nodes remain silent, which can provide nodes with a conflict-free transmission mode and high reliability of data transmission. On the basis of TDMA, more research proposals followed, by combining TDMA with frequency division multiple access (FDMA), code division multiple access (CDMA) and space division multiple access (SDMA), driving further development of TDMA to ensure conflict-free transmission of data streams in time, frequency, code and space. However, to achieve the conflict-free transmission mechanism of TDMA requires ensuring precise synchronization between individual nodes. Although microsecond-level synchronization accuracy is currently achieved in wireless networks, this imposes additional load overhead on the network and due to the poor scalability of TDMA, the cycle time of data stream transmission increases linearly when the number of nodes in the network increases, which in turn leads to high latency, thus it is not practical to use TDMA in dense wireless sensing networks.
In time division multiple access (TDMA), each node access the channel in a dedicated time slot of each frame while all other nodes remain silent, and thus providing the nodes a conflict-free MAC.
As a well-established method for providing deterministic communications, TDMA plays a key role in reducing channel access latency and has been applied in many wireless systems \cite{TDMA}.
The drawback of TDMA is that its implementation requires the precise synchronizations among all the nodes, which imposes additional overhead to the network.
     Besides, the frame length increases linearly with the number of nodes in the network, which introduces large delays in large-scale networks.
Also, the TDMA based networks are not very flexible to the changes in nodes, i.e., joining the network or leaving the network. Thus, it is difficult for TDMA to be used directly in scalable networks.

\subsubsection{OFDMA}
%The packets transmitted in the channel always consist of a certain number of symbols (depending on the number of bytes transmitted, and other, for example, PHY parameters of the modulation and coding scheme), which are used to control the overhead as well as the payload, respectively. Before WiFi 6 (802.11ax), data transmission was done in OFDM mode, by dividing the channel into several orthogonal sub-channels, converting high-speed data traffic into parallel low-speed sub-data traffic, modulated to transmit on each sub-channel, this way the sub-channel could transmit data without suffering from multi-path distortion of the same strength faced by single-carrier transmission. In an OFDM system, only one user can transmit on all subcarriers at any given time. To accommodate multiple users, OFDM needs to support multiple access through TDMA (separate slots), as shown in Fig. \ref{OFDM.eps}. The disadvantage of OFDM is that it does not take advantage of the fact that different users see the wireless channel in different ways. 802.11ax proposes the OFDMA operation mode, which improves efficiency and reduces queuing delay by dividing the resources of the entire channel into fixed-size time-frequency Resource Units (RU) and carrying data from users on each RU, thus enabling multiple users to transmit in parallel at the same time in each time slot without having to wait in line and compete with each other, as shown in Fig. \ref{OFDMA.eps}.

\begin{figure}
  \centering
  \includegraphics[width=3.7in]{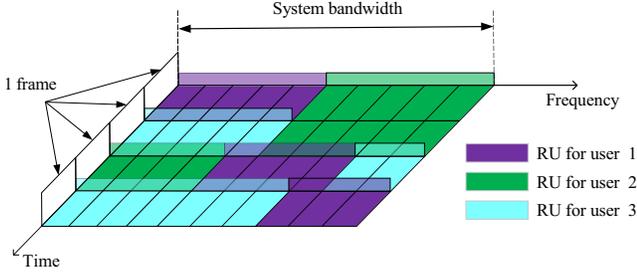}\\
  \caption{Resource structure in OFDMA.}\label{OFDMA.eps}
\end{figure}

Before the advent of Wi-Fi 6 (802.11ax), WLAN has been using the orthogonal frequency division multiplexing (OFDM) based modulations. %, as shown in Fig. \ref{OFDM.eps}.
    In OFDM, the frequency bandwidth is divided into several orthogonal sub-channels, so that a high-speed data traffic flow is separated into a set of parallel low-speed sub-data traffic flows (i.e., sub-flows).
By mapping the sub-flows to sub-channels, the frequency selective fading suffered  by single-carrier transmissions could be avoided.
    However, only one flow can be transmitted over each sub-channel, which reduces the spectrum efficiency.
To address this issue, 802.11ax proposes to use orthogonal frequency division multiple access (OFDMA), as shown in Fig. \ref{OFDMA.eps}.
    In OFDMA,  the frequency and time resources are divided into  time-frequency resource units (RU) of fixed-size, so that the data of different users are carried on RU resource blocks other than sub-channels.
Therefore, OFDMA enables higher flexility in resource allocation and improves reliability while reducing queuing delay. It is also noted that OFDMA is more suitable for downlink communications other than uplink transmissions.

\subsubsection{NOMA}
%Non-orthogonal multiple access (NOMA) allows multiple users to transmit in the same band, relying on the Successive Interference Cancellation (SIC) technique at the receiver to gradually eliminate interference to iteratively decode the signal (i.e., at the receiver side, the multiple access interference from successfully decoded packets in previous rounds can be removed from the received signal to improve the signal-to-noise ratio of the remaining packets, which is especially important for NOMA). The orthogonal multiple access techniques such as TDMA, FDMA, OFDMA, and others can achieve higher frequency efficiency when the number of users is less. However, high latency is caused by collisions when the number of users increases, while NOMA eliminates the random access phase when the number of users is high, effectively using interference cancellation for decoding at the base station and providing low latency performance, which is important for time-critical communications, hence NOMA is considered for 5G URLLC use cases.

Non-orthogonal multiple access (NOMA) use the power multiplexing technique which allow multiple users to transmit in the same frequency band simultaneously.
    By using the successive interference cancellation (SIC) method at the receiver side, the receiver decodes the data of different users successively.
Once the data of a user is decoded, its interference to remaining users would be subtracted from the received signal, so that the signal-to-noise ratio of the remaining signals can be increased.
It has been proved that the orthogonal multiple access (OMA) techniques (e.g. TDMA, OFDMA) performs slightly better than NOMA in terms of latency when the number of users is small, while NOMA performs better than OMA when the number of users is large \cite{H.C}. This is because when the number of users is small, each user can be allocated with a separated and orthogonal  piece of resource so that spectral efficiency is higher when transmitting through OMA; when the number of devices is large, there would be not enough resources to maintain the orthogonality among users. This leads to many unavoidable collisions and large latencies. On the contrary, NOMA allow a large number of devices to share the same radio resources and provide them with the desired latency.
    Another advantage of NOMA is that it does not require a separate authorization and random access phase, as devices can send data at any time.
%This is because under high traffic loads, random access contention causes inevitably high latency, while NOMA requires no separate authorization acquisition and random access phases, and allows users to share the same bandwidth resources to support a large number of devices with the required latency.
This is the scenario of interest for most IoT use cases, and for URLLC scenarios where a large number of users joins and leaves the network frequently (e.g., intelligent traffic systems and logistics monitoring systems).
    In fact, NOMA has been selected as a key technique for 5G URLLC use cases.
However, the implementation of NOMA technology still faces some difficulties. For example, the power multiplexing technology is not  mature; the non-orthogonal receiver is quite complex; interferences from other nodes; the channel states are needed to restore and remove;
 designing the SIC NOMA-receivers also depends on further advancement in signal processing chips.

\subsubsection{CSMA/TDMA based Hybrid MAC}
%The carrier listening multiple access (CSMA) introduced by IEEE 802.15.4 is a simple and effective conflict avoidance mechanism which is highly scalable compared to TDMA, does not require time synchronization, and exhibits higher channel utilization and lower latency than TDMA under low network load, as well as the disadvantages of high packet loss rate, sharp increase in transmission delay, and high node energy consumption during network congestion. Under the demand of wireless sensor network development, the hybrid CSMA/TDMA MAC protocol has emerged, which realizes the complementary advantages of CSMA and TDMA and ensures low-latency and high-reliability transmission of data while guaranteeing a small packet loss rate, which has become the main solution for many wireless sensing networks. However, in the CSMA and TDMA protocols, each node can effectively avoid conflicts through the fallback mechanism, but each device can access the spectrum at any time and cannot transmit time-critical traffic in real time, so it is not suitable for wireless networks with multiple priority traffic sharing.
The carrier sense multiple access with collision avoidance (CSMA/CA) protocol is a contest-based MAC protocol with high flexibility and scalability.
 However, the CSMA protocol is inefficient in terms of throughput due to the time wastage caused by the back-off process.
 On the contrary, the CSMA/TDMA hybrid MAC protocol improves determinacy by combining the advantages of CSMA and TDMA.
    Specifically, CSMA/TDMA does not require the frame synchronization among nodes.
        In case the traffic load is relatively light, the devices access the channel using the CSMA/CA scheme.
    In case the traffic load is relatively heavy,  the devices would send a slot request (indicating how many slots are needed) to a central coordinator, which allocates some reserved slots for the node.
        In doing so, higher channel utilization and lower latency can be guaranteed.
 In CSMA/TDMA, however, all nodes must perform low-power listening in all the time slots to monitor the possibly incoming data, which is challenging for most battery-powered sensor nodes \cite{N.A, W.W}.

\subsubsection{ARQ and HARQ}

\begin{figure}[htbp]
\centering

\subfigure[Average number (${T}_{\text{p}}$) of slots to deliver a packet versus the probability $p$ of successful packet transmission, in which ${p}_{f}=0.8$ and $C=5$ slots.]{
\begin{minipage}{0.47\textwidth}
\centering
\includegraphics[width=3.5in]{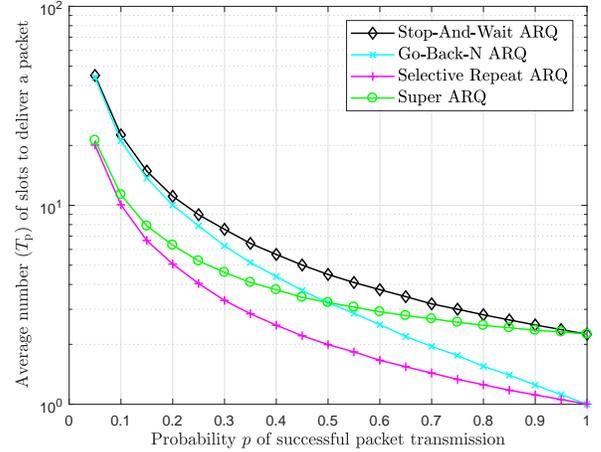}\label{ARQ_p_0.8}
%\caption{fig1}
\end{minipage}%
}%
\quad
\subfigure[Average number (${T}_{\text{p}}$) of slots to deliver a packet versus the probability  $p_f$ of successful feedback transmission, in which $p=0.4$ and $C=5$ slots.]{
\begin{minipage}{0.47\textwidth}
\centering
\includegraphics[width=3.5in]{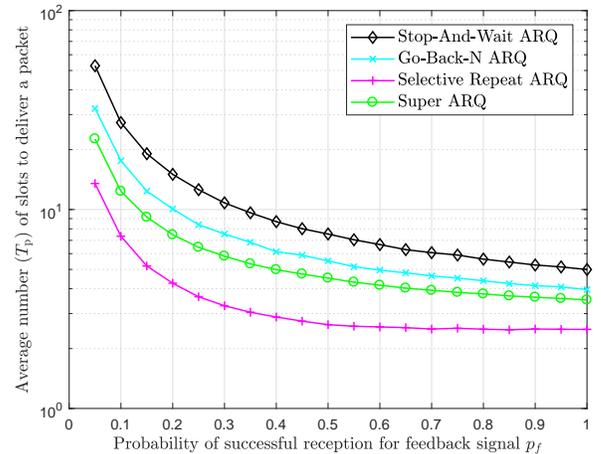}\label{ARQ_p0.4}
%\caption{fig2}
\end{minipage}%
}%

\centering
\caption{Change in the average number (${T}_{\text{p}}$) of slots to deliver a packet.}
\end{figure}

In wireless networks, the reliability of data transmissions is guaranteed by retransmission mechanisms.
    In the conventional automatic retransmission request (ARQ) mechanism (e.g., stop-and-wait ARQ), the receiver returns an ACK to the transmitter if the decoding of the received signal is successful and returns a NACK  to start a retransmission otherwise.
The continuous ARQ protocol allows the transmitter to send packets continuously without waiting for a confirming feedback, including the go-back-N ARQ protocol and the selective repeat ARQ protocol.
    In the go-back-N ARQ protocol, the receiver sends a feedback (NACK) only for unsuccessful decodings.
Upon receiving a NACK, the transmitter would retransmit the corresponding packet and all the following packets.
    In the selective repeat ARQ protocol, each packet is labeled with a timer, which indicates the number of allowed slots to wait for the feedback of the packet. If an ACK is not received within the period indicated by the timer, the transmitter will retransmit the packet.
On the contrary, in a super ARQ protocol, the transmitter keeps retransmitting a packet until its ACK packet is received.
    In each slot, we denote the probability of successfully transmitting a packet over the fading channel as $p$ and denote the probability of successfully transmitting an ACK/NACK over the feedback channel as ${p}_{f}$. We set the timer to $C$ slots.
It assumes that the latter packet is transmitted immediately after completing the transmission of the previous packet. Denote the number of slots required for a packet to complete its transmission as ${{T}_{\text{p}}}$, which includes the time to transmit and retransmit the packets, as well as the time to transmit the ACK and NACK. Specifically, ${{T}_{\text{p}}}$ can be expressed as
\begin{large}
\begin{align}
{{T}_{\text{p}}}={{T}_{\text{t}}}+\sum\limits_{i=1}^{M}{{{T}_{i}}},
\end{align}
\end{large}
in which ${{T}_{\text{t}}}$ is the number of slots to transmit and retransmit the packet, ${T}_{i}$ is the time to feedback an ACK or a NACK, and $M$ is the number of used ACKs and NACKs. It is clear that ${{T}_{\text{t}}}$ follows the geometric distribution with parameter $p$ and ${T}_{i}$ follows the geometric distribution with parameter ${p}_{f}$. In the stop-and-wait ARQ, the next packet cannot be started until the ACK of current packet. Note that the transmission and retransmission consume ${{T}_{\text{t}}}$ slots means that there are one slot to transmit the packet (which fails) and ${{T}_{\text{t}}}-1$ slots to retransmit the packet (only the last retransmission is successful). Thus, the receiver will feedback ${T}_{\text{t}}-1$ NACKs and one ACKs, i.e., $M={T}_{\text{t}}$. In the go-back-N ARQ, we have $M={T}_{\text{t}}-1$, since the receiver only feeds back the NACKs. In the super-ARQ, the transmitter keeps retransmitting a packet until it receives the feedback ACK. Thus, we only need to consider the delay generated by feedbacking ACKs, i.e., $M=1$. For the selective repeat ARQ, if the transmitter does not receive an ACK within the period defined by the timer of  $C$, the packet would be retransmitted, and thus ${T}_{\text{p}}$ can be expressed as
\begin{large}
\begin{align}
T_{\text{p}}^{i}
=\left\{ \begin{aligned}
  & T_{\text{t}}^{i},&& 1+T_{\text{f}}^{i}\le C \\
  & T_{\text{t}}^{i}+T_{\text{p}}^{i+1},&& 1+T_{\text{f}}^{i}>C, \\
\end{aligned} \right.
\end{align}
\begin{align}
{{T}_{\text{P}}}=T_{\text{p}}^{1},
\end{align}
\end{large}
in which $T_{\text{p}}^{i}$ is the number of time slots for the packet to complete its transmission by the $i$-th retransmission, $T_{\text{t}}^{i}$ is the number of time slots to deliver the packet during the $i$-th retransmission, $T_{\text{f}}^{i}$ is the number of time slots for feedbacking the ACK in the $i$-th retransmission. Note that $T_{\text{t}}^{i}$ and $T_{\text{f}}^{i}$ obey geometric distributions with parameters $p$ and ${p}_{f}$, respectively.
In Fig.\ref{ARQ_p_0.8} and Fig.\ref{ARQ_p0.4}, we compare the number of transmission slots to deliver a packet over a fading channel for the above mentioned protocols by simulation.
    From the Fig.\ref{ARQ_p_0.8} and Fig.\ref{ARQ_p0.4}, we observe that the transmission delay of the selective repetitive ARQ is the smallest.
However, the packets may not be received with correct order and the packet repetition probability is relatively large.
    In addition, the go-back-N ARQ protocol is a satisfying choice in case the probability of successful decoding is large, i.e., the channel condition is good.

%\begin{figure}
%  \centering
%  % Requires \usepackage{graphicx}
%  \includegraphics[width=3.5in]{ARQ_p_0.8.eps}\\
%  \caption{\replaced{Average number (${T}_{\text{p}}$) of slots to deliver a packet}{Average transmission delay of each block} versus the probability $p$ of successful packet transmission, in which ${p}_{f}=0.8$ and $C=5$ slots.}\label{ARQ_p_0.8}
%\end{figure}
%%Fig.\ref{ARQ_p_0.5} presents how the average number of slots per packet changes with the probability of successful packet reception $p$. From the figure we can see that as $p$ increases, the average number of slots required to confirm successful transmission per packet gradually decreases, and compared to other transmission schemes, the Selective Repeat ARQ scheme generates the lowest transmission delay. The Super ARQ scheme generates lower latency than the Go-Back-N ARQ when the downlink channel quality is poor, while on the contrary when the downlink channel quality is good.
%
%\begin{figure}
%  \centering
%  % Requires \usepackage{graphicx}
%  \includegraphics[width=3.5in]{ARQ_p-0.4.eps}\\
%  \caption{\replaced{Average number (${T}_{\text{p}}$) of slots to deliver a packet}{Average transmission delay of each block} versus the probability  $p_f$ of successful feedback transmission, in which $p=0.4$ and $C=5$ slots.}\label{ARQ_p0.4}
%\end{figure}

%
%%In Fig.\ref{ARQ_p0.4}, we presents how the average number of slots per packet changes when the probability of successful reception for feedback signal ${p}_{f}$ increases. It can be seen that the packet transmission delay gradually decreases as the uplink channel quality becomes better, and Selective Repeat ARQ shows the best performance in terms of latency. However, compared with other ARQ schemes, packets in the Selective Repeat ARQ scheme are not received in order at the receiving end, which may cause some packets to be dropped when they are received because they exceed the deadline, and the size of timer $C$ should be as low as possible in applications with high timeliness requirements, which may result in wasted resources and increased latency due to duplicate packet transmissions.

Another widely used retransmission protocol is the hybrid ARQ (HARQ), which combines the forward error correction coding (FEC) and ARQ.
    Specifically, the transmitter encodes each packet with a certain FEC coding scheme. When the receiver fails to decode a packet, it saves the received signal and feedbacks the transmitter a NACK to start a retransmission.
Upon receiving the retransmitted signal, the receiver decodes the packet based on an optimized combination of the retransmission and previous unsuccessful attempts other than the retransmission itself.
    Compared with other ARQ protocols, HARQ can efficiently reduce the probability of unsuccessful decoding, and thus increases the transmission rates while reduces the transmission delay.
However, the use of FEC coding requires additional overhead bits in each packet.
    Thus, a reasonable balance between transmission reliability and the transmission efficiency should be considered.

\subsubsection{TSCH}

Time synchronous channel hopping (TSCH) is a MAC layer scheduling scheme specified by IEEE 802. 15. 4 \cite{D.D, D-TSCH}. It is designed to provide high reliability channel access for low power IoT networks.
     Specifically, TSCH  combines the TDMA, frequency hopping, and the HARQ retransmission mechanism.
When a packet transmission fails, it would be retransmitted in the next time slot on a different frequency with the HARQ protocol, which provides a higher probability of success than using the same frequency again.
    Thus, TSCH exploits the frequency diversity and time diversity effectively. More, TSCH can also be combined with path diversity techniques (e.g., packet replication) that allows packets to be transmitted over multiple independent paths.
In doing so, TSCH reduces the packet loss and the end-to-end delay of the network effectively.
    Different from CSMA/CA, the low-power feature of TSCH allows routing nodes to reduce power consumption and ensures better latency control.
In addition, TSCH provides a resource reservation mechanism that allows slots to be reserved in advance for time-critical data traffic, and thus ensures better determinacy.

\subsubsection{Black Bursts}

In case there are both high-priority real-time traffic and low-priority non-real-time traffic in a network, Sobrinho and Krishnakumar proposed an Ethernet quality of
service using black bursts (EQuB) mechanism to provide bounded delay for high-priority traffic \cite{EQuB1}.
    Specifically, the low-priority nodes tries to access the channel by using the binary exponential back-off algorithm.
In case a high-priority node participates the channel contention, it transmits a jamming sequence (i.e., a black burst) with a pre-defined length to stop other node from accessing the channel.
    The use of black bursting requires mandatory modifications to the MAC and physical layers, which inevitably changes the use of components off-the-shelf (COTS).
The black bursting mechanism has been proved to be effective in providing high QoS and has been applied in various real-time wireless applications.% \cite{Suthaputchakun, Bi, P}.

\subsubsection{Transmission Protections}
%Since the real-time priority of RT traffic is much higher than that of RL traffic and BE traffic, most of the research schemes focus on improving the real-time performance of RT traffic without taking into account the transmission of low-priority traffic, especially under the preemption mechanism, which cannot guarantee the Worst-Case end-to-end Delay (WCD) of low-priority traffic. Therefore, it is necessary to guarantee the WCD of individual data traffic by some other protected transport mechanisms.

Most existing real-time transmission schemes focus on high-priority real-time traffic without considering the influence on those low-priority traffic.
    For example, the preemption mechanism cannot guarantee the worst-case end-to-end delay (WCD) for the low-priority traffic.
Therefore, it is necessary to guarantee the WCD of individual data traffic through adjustable protected transmission mechanisms, as detailed in the following.

\begin{itemize}
\item %Power adjustment: The probability of a packet being successfully transmitted in a time slot can be adjusted by adjusting the amount of transmission power at the sender side of the packet, so that the packet can be made to complete transmission before the deadline by increasing the transmission power when some packets are close to the deadline, and if necessary, the transmission power can be increased to ensure that the packet is successfully transmitted in a single time slot. By doing so, the WCD of low-priority traffic can be guaranteed without degrading the performance of high-priority traffic.
    Power adjustment:  Note that by adjusting the instant transmit power, the probability of successful packet transmission can be changed on demand.
        When the (re)transmission of some packets is close to its deadline, we can increase its transmit power so that the transmission
    packets could be completed before the deadline.
\item
    Bandwidth allocation: To meet the requirement of real-time applications in wireless networks, the limited bandwidth needs to be allocated for high-priority traffic flows first and allocated for low-priority traffic flows when there is spare bandwidth.
    Among the TSN-related protocols, IEEE 802.1 Qci defines a flow filter that can block all future frames of a particular traffic if unexpected/excessive bandwidth usage is detected;
    IEEE 802. 1Qcc improves the stream registration protocol (SRP) by automatically reducing the bandwidth and frame size of flows if there is no sufficient bandwidth resources for its reservation.
    Bandwidth resources can also be allocated and optimized according to the probability distribution of traffic flows with different priority levels, so that when bandwidth for high-priority traffic is insufficient, the bandwidth allocation between the low-priority flows and the high-priority traffic flows can be adjusted. However, the throughput achieved through bandwidth allocation is still limited, as the available sub-6-GHz spectrum is insufficient to meet the extreme bandwidth demands of data-intensive IIoT applications. Therefore, a beyond sub-6-GHz band is required for IIoT.
\item The 1:1 and 1+1 protection: %In order to guarantee the reliability of data transmission in the network, it is necessary to protect the data during transmission in the channel, which is divided into two protection modes, 1:1 and 1+1. The "1:1" protection mode has two transmission channels, the primary channel and the backup channel. During normal operation, the data traffic is transmitted in the primary channel and the backup channel can transmit low priority traffic. In the event of a failure on the primary channel, the sender and receiver negotiate and make a decision to switch the data traffic from the primary channel to the backup channel at the same time. The "1+1" protection mode allows packets to be transmitted in both channels simultaneously and detects the data transmission quality of both channels at the receiving end, receiving packets from the channel with the higher transmission quality.
    The ``1:1" protection mode uses two transmission channels, i.e., a primary channel and a backup channel.
        During normal operations, data traffic is transmitted through the primary channel while the standby backup channel can be shared with other low-priority traffic \cite{CLC-2021}.
    In  case  a transmission failure occurs on the primary channel, the transmitter and receiver would negotiate and make a decision to switch the transmission from the primary channel to the backup channel.
        The ``1+1" protection mode allows the traffic flow to be transmitted on the both channels at the same time.
        The receiver monitors the quality of the channels and decodes packets from the one with a higher reliability.
\item Dynamic priority assignment: Variable priority transmission allows the transmitter to increase the priority of a low-priority traffic flow when its packets are approaching their deadlines.
        This dynamic priority assignment can reduce the end-to-end delay of those initially assigned low-priority traffic effectively.
\end{itemize}
%In addition, the use of transmission protection can have some negative effects, such as increasing power will cause interference to other users on adjacent bands, and providing protected mode for data will waste some bandwidth resources. However, in real-time applications, it is critical to reduce end-to-end latency and jitter, so transmission protection is also an important way to improve real-time performance.

\subsubsection{WTSN}
Recent advances in 5G and IEEE 802.11 based wireless accessing technologies have boosted significant interest to wireless time-sensitive networking (WTSN). Note that some MAC layer protocols associated with TSN can effectively reduce end-to-end latency and improve transmission reliability.
    For example, IEEE 802. 1Qbv meets the strict delay bounds of high-priority traffic through the gate control list (GCL) mechanism and reduces the interference between traffic flows with different real-time priorities by separating them with protection bands;
IEEE 802. 1CB increases transmission reliability by transmitting redundant packets over independent paths and eliminating some duplicated packets at or near the receiver;
%    There are also technologies that provide functions similar to TSN through antennas (e.g., 5G-NR meets most industrial case requirements through URLLC mode), and extending TSN functions and standards to wireless networks is the subject of research and standardization activities.
With some proper modifications, some TSN functionalities can also work well over wireless media and wireless networks. In fact, a portion of the 802.1 TSN standards (e.g., 802.1AS (time synchronization), 802.1Qav (credit-based traffic shaping), 802.1Qbv (time-aware scheduling), and 802.1Qca (path control and reservation)) have been shown to be applicable to wireless standards (e.g., 802.11 and 5G) \cite{D.C}. In addition, 802.11be (defined by TGbe) have provided several new features on better integration with 802.1 TSN standards. For example,  the newly defined multi-link/channel operation reduces congestion by separating traffic of different priority levels; the multi-AP functionality enables multi-AP to improve link reliability by exploiting spatial diversity gain.

\subsection{Physical Layer Techniques}
%The primary task of the physical layer is to deliver messages over the channel in a reliable manner as fast as possible. Due to coexistence with other wireless systems, the transmission of packets suffers from band interference, which generates uncertain delays and jitter, and the delay and reliability performance cannot be improved simultaneously when the packets are short. Therefore, proper allocation of bandwidth, low latency and reliable transmission of short packets is a major challenge for the PHY layer.

In order to connect smart devices (both fixed and mobile) to distributed industrial data centers, the 5G-based IIoT networks need to provide reliable wireless accessing services. Specifically, a large amount of data needs to be transmitted at a very high rate (up to 40 Gb/s for stationary devices and 5 Gb/s for mobile devices), while centimeter-level positioning accuracy needs to be guaranteed under multi-path conditions \cite{A.V}. This has driven extensive research in advanced signal processing methods at the physical layer, aiming at transmitting information over the channels with high reliability and high data rates.

%\subsubsection{Bandwidth Allocation}
%To meet the requirement of real-time applications in wireless networks, the limited bandwidth needs to be allocated for high-priority traffic flows first and allocated for low-priority traffic flows when there is spare bandwidth.
%    Among the TSN-related protocols, IEEE 802.1 Qci defines a flow filter that can block all future frames of a particular traffic if unexpected/excessive bandwidth usage is detected;
%    IEEE 802. 1Qcc improves the stream registration protocol (SRP) by reducing the reserved bandwidth and frame size of flows if there is no sufficient bandwidth resources are available.
%    Bandwidth resources can also be allocated and optimized according to the probability distribution of traffic flows with different priority levels, so that when bandwidth for high-priority traffic are insufficient, the bandwidth allocation between the low-priority flows and the high-priority traffic flows can be adjusted.
\subsubsection{mmWave Communications}
With advantages in larger bandwidth, lower interference, and more abundant spectrum, the mmWave communication is widely considered as a prospective solution to future wireless networks with demand on increasingly higher data rates and lower latency \cite{S.S}.
    For example, the capacity of mmWave-based 5G system is ten times larger than traditional systems, and thus is applicable for future IoT-based factory deployments \cite{B.M}.
Note that the mmWave antennas are much smaller than those traditional antennas using lower frequency bands, since  the minimum length of a antenna increases linearly with the wave-length.
    Thus, many mmWave antennas can be placed in small size devices (e.g., smart phones) as mmWave-mMIMO antenna arrays to combat path loss \cite{J.Z}.
Modern on-chip mmWave-mMIMO arrays have also proved to be cost effective and power efficient for industrial IoT devices \cite{W.Hong}.
    Moreover, by using the unlicensed millimeter wave spectrum and the TDMA scheduling in  ultra-dense radio access networks (RAN),  smaller delays and jitters can be realized at a near-minimum transmission power in Wi-Fi systems \cite{T.H.S}, which provides a feasible solution for time-critical industrial wireless communications.
%However, there are still many issues to be solved to realize mmWave communications, such as unavailability of multi-path components due to higher energy scattering, short transmission distances caused by high path loss and high LoS blocking probability resulting from random channel conditions. Finally, mmWave communications also presents some challenges in the MAC layer design, for example, the need for high-directional antennas, relatively short LoS paths and beam formation \cite{H.S}.
Nevertheless, there are still some issues that need to be considered in applying mmWave communications in time-critical industrial IoT networks.
\begin{itemize}
  \item In case the used frequency bandwidth is very large (e.g., over 1 GHz), the corresponding A/D converter and signal processing unit would be much more expensive and less energy-efficient \cite{T.S.R}.
  \item The path-loss attenuation is large for  mmWave communications.
            In order to reduce interference and attenuation,  it is critical to perform directional beamforming and transmit on LoS path in mm-Wave communications \cite{C.C.G}.
  \item While beamforming provides higher antenna gain by using narrower beams, the corresponding multiplexing gain is smaller due to less multi-path components.
                In fact, parallel multi-path transmissions between multiple transmitter-receiver pairs are crucial to increase the reliability and data rate, in terms of diversity and multiplexing gain, respectively.
            Therefore, a balance between array gain (beamforming) and spatial multiplexing gain  should be considered \cite{A.M-2021}.
  \item The penetration capability of mm-Wave signals is poor due to its high frequency.
            Furthermore, since the scattering level is high and the objects are mobile in industrial environments, the probability of LoS blocking is high.
            To this end, we need to provide effective beam management (e.g. tracking, serving) in mmWave communications \cite{M.G}.
\end{itemize}

\subsubsection{Modulation and Coding Schemes}
%Modulation and coding scheme is an important aspect that causes latency and reliability. With different block lengths, different levels of reliability and latency need to be provided flexibly. On the one hand, for short packets commonly used in applications with high real-time performance requirements, the code block length will be limited in order to achieve low latency, but as the code block length is reduced, the coding gain of the channel gradually decreases and these codes will suffer significant performance loss and reduce the reliability of data transmission. On the other hand, using high-complexity decoding algorithms, reducing the code rate to increase channel coding redundancy, and employing hybrid automatic retransmission request techniques will provide higher reliability for data transmission, but all will increase latency inevitably.

In industrial IoT networks, the packets are usually short, with a payload size of a few bytes.
    Thus, the controlling overhead (e.g., synchronization information, channel status) is comparable to the payload or even higher \cite{M.L}.
        On the one hand, the code block length is very limited in industrial applications, so that the coding gain is also limited, which reduces the reliability of transmissions.
On the other hand, by using high complexity decoding algorithms and the HARQ scheme, although higher reliability of transmissions can be guaranteed, and the corresponding latency will inevitably be increased.
    Therefore, efficient modulation and coding schemes are essential to achieve deterministic communications.
\begin{itemize}
\item %Fixed-rate channel code: Fixed-rate channel codes for low latency and high reliability requirements have achieved significant results, such as BCH codes \cite{S}, tail-biting convolutional codes \cite{L}, Low Density Parity Check (LDPC) codes \cite{D}, and polar codes \cite{E} are considered as candidate channel codes for URLLC. The minimum distance of BCH code is larger to avoid performance degradation in case of low Block Error Rate (BLER), but the code block length and message length cannot be chosen arbitrarily. The tail-biting convolutional codes solve the rate loss problem due to zero-tailed termination of short block lengths, as well as having a high decoding complexity. The coding efficiency of LDPC codes is close to the Shannon limit, but the decoding complexity of LDPC codes increases with flexibility and has a certain lower error limit at a certain rate and block length. And polar codes have better performance than LDPC codes in terms of short block length and low code rate, as well as exhibit no error lower bound, but have higher decoding complexity at larger block lengths.

    Fixed-rate channel codes: The BCH codes, tail-biting convolutional codes, low density parity check (LDPC) codes, and polar codes are considered as strong candidates for URLLC. %\cite{S} \cite{L} \cite{D} \cite{E}
        Specifically, although the minimum distance of BCH codes is relatively large and the block error rate is smaller, the code block length cannot be chosen arbitrarily and is not so flexible;
    while the tail-biting convolutional codes solve the rate loss problem due to its zero-tailed termination in short block lengths, the encoding/decoding complexity is high;
        the coding rate of LDPC codes can closely approach the Shannon Limit, but the decoding complexity increases with flexibility;
    while the polar codes perform better than LDPC codes in the short block length regime and are not limited by any error lower bounds, the decoding complexity is higher as the block length becomes larger.

%        Specifically, the minimum distance of BCH codes is relatively  large and the  block error rate smaller, however, the code block length cannot be chosen arbitrarily and is not so flexible; the tail-biting convolutional codes solve the rate loss problem due to its zero-tailed termination in short block lengths while the encoding/decoding complexity is high; the coding rate of LDPC approaches the Shannon Limit but the decoding complexity increases with flexibility; while polar codes perform better than LDPC codes in the short block length regime and are not limited by any error lower bounds, the decoding complexity is higher as the block length becomes larger.

\item %Rate adaptive scheme: The coding scheme of data packets is not fixed, and choosing a suitable coding scheme under a wide range of channel conditions can effectively reduce the delay and improve the reliability of data transmission. The rate adaptive scheme can choose a suitable coding scheme according to the real-time state of the channel, i.e., before each transmission, the transmitter sends a guide signal to the receiver so that the receiver can estimate the channel state, and the receiver feeds back the Channel Quality Indicator (CQI), and based on this CQI, the transmitter selects the best combination of fixed-rate channel code and modulation scheme from a predefined set, and if information cannot be recovered by the receiver it will be retransmitted according to HARQ. In reliable communication, feedback is necessary, but retransmission should be avoided in time-critical applications, especially when packets are short and cannot be retransmitted with HARQ because of the large delay, so use of a rate-adaptive scheme needs to consider channel codes that are suitable for short block lengths and low BLERs.
    Rate adaptive schemes: Modern channel estimation techniques are providing transmitters with more and more accurate CSI, with which the transmitter can determine its coding rate adaptively so that the desired block error rate can be guaranteed.
        The main challenge for using rate adaptive schemes in IoT networks is that the overhead is relatively large due to the limited block length.

\item %Rateless code: %In the physical layer, the delay caused by the channel estimation time is not negligible, so to achieve the low delay requirement, transmission schemes that are not dependent on CQI should be considered, and the three components of data (frequency guide, control and data) need to be combined in order to reduce the delay. With the characteristics of feedback-free link and the ability to dynamically determine the code rate according to the channel state during transmission, rateless codes do not require CSI at the transmitter side, avoiding CSI feedback and retransmission, which is a promising solution to meet the low latency and high reliability requirements of URLLC. Conventional rateless codes, such as LT codes \cite{M} and Raptor codes \cite{A}, have been shown to approach infinite block-length channel capacity over binary erasure channels, but the achievable rate performance of rateless codes degrades significantly in low signal-to-noise regions and therefore cannot be used alone over physical channels in URLLC.
    Rateless code: A rateless code encodes the information to be transmitted to a large number of short packets without requiring the CSI of the channel.
        These packets are then transmitted to the receiver without requiring feedback for each packet.
    Once the receiver has successfully received a certain number of these packets, the original information can be decoded successfully.
        %Note that the process of acquiring CSI at the transmitter and feeding the transmitter with a NACK is estimated to take 8--10 ms.
        Thus, the coding rate of rateless codes is adaptive to the channel automatically.
    This feedback-free  property makes rateless codes a prospective scheme to meet the low latency and high reliability requirements of URLLC applications.
        Conventional rateless code, such as LT codes  and Raptor codes \cite{A}, have also been shown to approach infinite block length channel capacity over binary erasure channels.
    In the low SNR regime, however, the achievable rate of rateless codes degrades significantly, so that they have to be used jointly with some channel codes.

\item %Analog fountain code: Analog Fountain Code (AFC) is a code with high throughput at arbitrary code rate and linear complexity, with a capacity close to that of rateless codes and suitable for a wide range of SNR. It has the advantages of rateless codes while solving the shortcomings of traditional rateless codes with significant performance degradation in low SNR areas, and provides a great solution for the design of the PHY layer of URLLC.
    Analog fountain code: The analog fountain code (AFC) extends traditional rateless codes by coding over the Euclidean space other than the Hamming space.
        Thus, AFC can be readily used for adaptive modulations.
    It was also shown that the capacity of AFC is close to that of rateless codes for a wide range of SNR and the coding complexity is linear with arbitrary code rate.
        These features enables AFC a prospective solution for the physical layer design of URLLC \cite{M.S}.
\end{itemize}
%The combination of a rate adaptive scheme with analog fountain codes can be effective in reducing the delay and improving the reliability of the transmission. However, the receiver is still required to know the CQI in order to decode the information and the transmitter needs to insert the guide symbols into each transmitted block (without the transmitter knowing the information about the channel state), which may cause severe performance loss, especially when the block length is short.

 Coding schemes that are adaptive for a wide range of channel conditions could reduce end-to-end delay and improve transmission reliability effectively.
     For rate adaptive schemes (even if rateless codes or AFC is used), the receiver still requires the CSI to decode the received packets, so that the transmitter has to insert pilot symbols in each block, which may lead to serious performance loss.
 The above mentioned self-adaptive channel codes also have some challenges and problems. For example, the decoding delay of AFC and that of other URLLC candidates (such as LDPC codes and polar codes) are comparable.

\subsection{An End-to-End Perspective}
\subsubsection{Time-triggered Mechanism}
In time-triggered mechanisms, the activities are initiated periodically, which reduces the randomness in the system \textit{in the first place} \cite{H.G}.
    On the contrary, traditional Ethernet uses an event-triggered mechanism, in which end systems/devices can access the network at any time.
Since the events arrive at these end systems randomly and their communications are carried out by the same medium, the accumulation of delays and jitter are inevitable and relatively large.
    Therefore, it is easier to provide deterministic communications by replacing event-triggered networks with time-triggered networks (e.g., TTP, TTCAN, TTE, FlexRay).

\subsubsection{Time Synchronization}
In industrial communications networks, multiple traffic flows of different priorities share the communications channels to improve efficiency and reduce costs.
     It is necessary to ensure the time synchronization among these traffic flows.
In wire-line networks, IEEE 802.1AS provides  bounded delay and extremely low delay variation for TSN applications through precise time synchronization for all the nodes, which is a fundamental requirement for implementing other standard functions.
    In wireless networks, it was shown that precise time synchronization and precise time protocol (PTP) applications can be supported by using hardware time-stamping in an open source testbed called \textit{openWIFI} \cite{wireless-sync}.
    %For example, IEEE 802.Qbv defines the protection band mechanism through time synchronization and time division multiplexing (TDM).
%Ensuring accurate synchronization of all nodes in the network requires the use of a large amount of bandwidth, i.e., imposes a significant overhead on the network, and may affect the overall scheduling mechanism of the nodes in case of synchronization errors.

\subsubsection{Traffic Shaping}
In case that some devices generate burst of the traffic over the network, traffic shaping \textit{at the sources} should be used to smooth the bursts according the available bandwidth.
    For example, IEEE 802.1Qcr reshapes the flows based on their priorities at each hop of transmission using an asynchronous traffic shaper (ATS); IEEE 802.1Qbv provides a time-aware shaper (TAS) through the GCL and a credit-based shaper (CBS) to prevent low-priority traffic shortages.
For wireless networks, traffic shaping and scheduling is even more important.
    By further considering the randomness of fading channels, extending ATS and TAS to wireless networks is necessary and feasible.

\subsubsection{Scheduling Transmissions and Decision-Makings}
\begin{figure}
  \centering
  \includegraphics[width=3.5in]{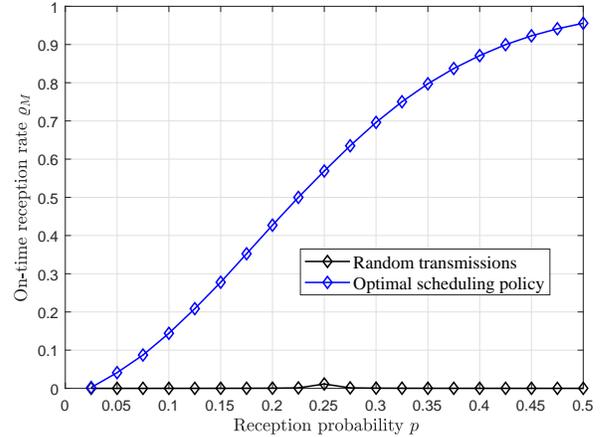}\\
  \caption{On-time reception rate versus the probability of success decoding of the fading channel, in which ${T}_{\text{tgt}}=4$ and $\delta=2$.}\label{on_time.eps}
\end{figure}

In addition to providing reliable transmission links, optimizing the desired decision epochs \textit{at the destinations}, i.e., the time to use the received information, is also crucial to deterministic networking.
    As is shown in \cite{AuD}, making decisions periodically leads to the smallest average age upon decisions (AuD), i.e., the average age of received update when they are used for decisions.
Moreover, for a sequence of periodic decision epochs, we can maximize the probability for the packets to be received at their respective decision epochs (i.e., on-time) by scheduling the \textit{transmission of packets} such as repeating, delaying and dropping the packets \cite{On-time-LY}.
    Suppose a sequence packets are transmitted to the receiver through a Rayleigh fading channel, the pre-defined reception slot interval of adjacent packets is ${T}_{\text{tgt}}$, and the $m$-th packet is expected to be received at the $m{T}_{\text{tgt}}$-th slot with a deviation no larger than $\delta$ slots.
We present how the on-time reception rate changes when the reception probability $p$ increases in Fig. \ref{on_time.eps}, in which the on-time reception rate is obtained by the Algorithm 2 in \cite{On-time-LY}.
From Fig. \ref{on_time.eps}, we observe that by optimally scheduling the packets, the on-time reception rate approaches 100\% as the probability of successful decoding from each transmission approaches $p=0.5$, which indicates a very unreliable channel.
    On the contrary, if no packet scheduling is used (i.e., using the random transmission scheme), the on-time reception rate is close to zero, regardless of $p$.

\section{Conclusion}\label{Conclusion}

This article analyzes the possibility to provide deterministic communications with wireless networks.
    On the one hand, traditional wireless channels using omni-directional antennae are highly unreliable while industrial IoT networks raised strict real-time requirements for information transmissions.
On the other hand, recent progress in wire-line deterministic networking technologies provided valuable benchmarks or references for wireless communications.
    By optimally choosing the MAC layer techniques (e.g., TDMA, packet scheduling, and transmission protection) and physical layer techniques (e.g., OFDMA and AFC),  it is believed that wireless deterministic networks will come to us in the near future.
It is also expected that when some machine learning method such as the Q-learning, multi-agent reinforce  learning, or the federated learning is used, the system design would be more flexible for deterministic communications scenarios.

\end{document}